\documentstyle[preprint,aps]{revtex}

\tighten

\begin{document}

\def\eps{\varepsilon}
\def\klein#1{{\mbox{\scriptsize #1}}}
\def\pzero{{\phantom{0}}}

\draft
\title{Crossover of magnetoconductance autocorrelation \protect\\
	for a ballistic chaotic quantum dot}

\author{K. Frahm}

\address{Service de Physique de l'\'Etat condens\'e.
	CEA Saclay, 91191 Gif-sur-Yvette, France.}

\date{\today}

\maketitle

\begin{abstract}
The autocorrelation function
$C_{\varphi,\eps}(\Delta\varphi,\,\Delta \eps)=$ $\langle
\delta g(\varphi,\,\eps)\,
\delta g(\varphi+\Delta\varphi,\,\eps+\Delta \eps)\rangle$ ($\varphi$
and $\eps$ are rescaled magnetic flux and energy)
for the magnetoconductance of a ballistic chaotic quantum dot is
calculated in the framework of the supersymmetric non-linear $\sigma$-model.
The Hamiltonian of the quantum dot is modelled by a Gaussian random
matrix. The particular form of the symmetry breaking matrix is
found to be relevant for the autocorrelation function but not for
the average conductance.
Our results are valid for the complete crossover from orthogonal to unitary
symmetry and their relation with semiclassical theory and
an $S$-matrix Brownian motion ensemble is discussed.
\end{abstract}


\narrowtext

GaAs/Al$_x$Ga$_{1-x}$As heterostructures provide useful
experimental realizations
\cite{marcus,keller,chang} of two-dimensional ballistic cavities
known as quantum dots. Measuring the conductance
of a quantum dot connected to electron reservoirs, one can
study the quantum behavior of classically chaotic billiards.
Apart from the channel number, the
magnetic field $B$ and the gate voltage $E$ are the two important
adjustable parameters. In experiments, the sample dependent conductance
is measured as a function of these parameters.
Theoretical approaches
\cite{bluemel1,jalabert3,jalabert4,frahm1,rau1,pluhar}
consider the weak localization peak (by calculating
the average conductance as a function of magnetic field) and
the conductance autocorrelation with respect to magnetic field
and gate voltage. Applying an ergodicity argument, one may compare
these quantities with experiment.

Chaotic ballistic systems were treated numerically and semiclassically
in \cite{bluemel1,jalabert3,jalabert4},
the latter predicting an algebraic decay of the correlation functions and the
weak localization peak.
A random matrix approach for the unitary scattering matrix
in terms of the circular Brownian motion ensemble \cite{frahm1,rau1} yields
an exponential decay of these quantities
as a function of a fictitious Brownian motion time $t$.
It is not obvious how to relate these results with the semiclassical
theory. Furthermore, it is not entirely clear if the problematic
diagonal approximation in the Gutzwiller trace formula, used in
Refs. \cite{jalabert3,jalabert4}, plays an important role here.

Recently, the supersymmetric non-linear $\sigma$-model has been applied
\cite{pluhar} to quantum dots to calculate the weak localization peak,
describing the crossover from orthogonal to unitary symmetry
for the average conductance.
In the unitary symmetry class where the magnetic field is sufficiently
strong to completely break the time reversal invariance,
the conductance autocorrelation function has been calculated
\cite{efetov2} in the limit of many channels.

Corresponding perturbative treatments of the $\sigma$-model
which are predominantly concerned with the metallic diffusive case
can be found in Ref. \cite{amg} for the magnetic field correlations and
in Ref. \cite{altland2} for the energy correlations, respectively.

The purpose of this letter is to generalize the treatment of Ref.
\cite{efetov2} to the complete crossover from orthogonal
to unitary symmetry. In addition, we point out that the precise form
of the symmetry breaking matrix is relevant for the autocorrelation
function but not for the weak localization peak. This observation is very
important to understand the Brownian motion results \cite{frahm1,rau1}.

We are interested in the generic chaotic features of the quantum dot
and model the Hamiltonian as a Gaussian random matrix of the form
\begin{equation}
\label{hamiltonian}
H_\alpha=H_\klein{GOE}^{(1)}+\alpha (\kappa H_\klein{GOE}^{(2)}+iA)
\end{equation}
where $H_\klein{GOE}^{(1,2)}$ are (independent) real symmetric
$N\times N$-random matrices given by the Gaussian orthogonal ensemble
\cite{mehta} with the variance $\lambda^2/N$ of the nondiagonal elements
and $A$ is a real antisymmetric
random matrix whose independent elements are normal distributed with
zero mean and variance $\lambda^2/N$. The parameter $\alpha$ (which is
proportional to the magnetic flux penetrating the quantum dot)
describes the strength of the symmetry breaking.
Concerning the parameter $\kappa$, which distinguishes between
different types of symmetry breaking,
we are mostly interested in the particular cases of a
purely imaginary antisymmetric symmetry breaking ($\kappa=0$)
and a hermitian symmetry breaking ($\kappa=1$). The first case yields
the well known Pandey-Mehta Hamiltonian \cite{pandey1} which is believed
to correspond to the application of a small magnetic field
\cite{bohigas5,frahm1}. The other case is rather
directly related \cite{frahm1} to the circular Brownian motion ensemble.

In the large $N$ limit, $\lambda$ is expressed in terms of
the average level spacing $\Delta_0$ (at energy $E=0$) \cite{mehta} by
$\lambda=N\,\Delta_0/\pi$ and the proper
crossover-parameter \cite{pandey1} to describe the symmetry
breaking is just $\sqrt{N}\alpha$.
Keeping this quantity finite as $N\to\infty$
the value of $\kappa$
does not affect $S$-matrix averages because the additional
contribution of $H_\klein{GOE}^{(2)}$
can be taken into account by an infinitesimal rescaling of the variance of
$H_\klein{GOE}^{(1)}$. On the other hand, correlations between different
values of $\alpha$ depend on the choice of $\kappa$.

We follow the treatment of Refs. \cite{pluhar,verbaarschot} and use for
the $M\times M$-scattering matrix the expression
$S=1-2\pi i W^\dagger(E-H_\alpha+i\pi WW^\dagger)^{-1}W$
where $W$ is a $N\times M$-matrix describing the coupling of the
scattering channels with the states of the quantum dot.
As in Ref. \cite{pluhar}, we use the choice of $W$ that
corresponds to the ideal coupling of the wires characterized
\cite{pluhar,verbaarschot} by a vanishing average $\langle S\rangle=0$
and leading to equivalent scattering channels. Then, the distribution of the
scattering matrix (for $\alpha=0$) is \cite{lewenkopf,brouwer} equivalent
to the circular orthogonal ensemble \cite{dyson1}.
We assume the dimension $M$ of the $S$-matrix to be even and
express the conductance by the Landauer formula
$g=\mbox{Tr}(S_{12}^\dagger S_{12}^{\phantom{\dagger}})$
where $S_{12}$ is the $M/2$-dimensional $(1,2)$-block of $S$.

We apply the supersymmetric approach as described in Refs.
\cite{verbaarschot,efetov1,altland1,iida1} to calculate the conductance
autocorrelation function at two different energies $E\pm\Delta E/2$
and two different values $\alpha\pm\Delta\alpha/2$
of the symmetry breaking parameter. We assume that in the large $N$-limit
the quantities $N\alpha^2$, $\ N\Delta\alpha^2$,
$\ E/\Delta_0$, and $\Delta E/\Delta_0$ remain finite.
The product of the two conductance contributions is expressed
as a Gaussian integral over a $16N$-dimensional supervector with
the $16N$-dimensional supermatrix
\begin{equation}
\label{h_super_def}
{\cal H} = \Lambda\biggl(E+\frac{\Delta E}{2}\Sigma_3
-\Bigl(H_\klein{GOE}^{(1)}+
(\alpha+\frac{\Delta\alpha}{2}\Sigma_3)\,
(\kappa H_\klein{GOE}^{(2)}+iA\tau_3)\Bigr)
+i\pi\Lambda WW^\dagger\biggr)\quad.
\end{equation}
We adopt here the notational conventions of Ref. \cite{altland1}.
The supermatrices $\Lambda$, $\tau_3$ and $\Sigma_3$ are diagonal matrices
with an equal number of diagonal entries $+1$ and $-1$, respectively,
defining
different types of gradings. $\Lambda$ describes the grading imposed
by the advanced and retarded Green's functions and the $\tau_3$-grading
is determined by the time reversal transformation. $\Sigma_3$
corresponds to the additional grading for the two conductance
contributions at different magnetic fields and energies.
As in \cite{altland1} we denote by $L_g$ the supermatrix describing
the decomposition in bosonic and fermionic sub-blocks.

Performing the usual steps \cite{verbaarschot,efetov1,altland1,iida1}
and using the rescaled quantities $\varphi=(8N\alpha^2/(M+1))^{1/2}$
and $\eps=2\pi E/[(M+1)\Delta_0]$ for the symmetry breaking parameter
and the energy, respectively, we arrive at the $Q$-integral
\begin{eqnarray}
\label{sigma_int1}
&&\left\langle {\textstyle g(\varphi-\frac{1}{2}\Delta\varphi,\,
\eps-\frac{1}{2}\Delta\eps)\,
g(\varphi+\frac{1}{2}\Delta\varphi,\,
\eps+\frac{1}{2}\Delta\eps)}\right\rangle\\
\nonumber
&&\qquad=\left(\frac{M^2}{32}\right)^2\int dQ
\prod_{\nu=1,2}
\left[\mbox{Str}((1+Q\Lambda)^{-1}Q\Lambda J_\nu)\right]^2
\ e^{-{\cal L}(Q)}
\end{eqnarray}
with the action
\begin{equation}
\label{act_def}
{\cal L}(Q)=\frac{M}{2}\mbox{Str}\,\ln(1+Q\Lambda)
+\frac{M+1}{32}\biggl(-4i\,\Delta\eps\,\mbox{Str}\left(Q\Sigma_3\right)+
\sum_{\nu=1,2}\mbox{Str}\left((Q\gamma_\nu)^2\right)\biggr)
\end{equation}
and the matrices
$\gamma_1=\Bigl(\varphi+\frac{1}{2}\Delta\varphi \Sigma_3\Bigr)\tau_3$, \
$\gamma_2=\frac{\kappa}{2}\Delta\varphi \Sigma_3$, \
$J_{1,2}=(1+L_g)(1\pm\Sigma_3)\,\sigma_1(\Lambda)/4$ \
($\sigma_1(\Lambda)$ is the first Pauli matrix in the $\Lambda$-grading).
The integration variable $Q$ is a $16\times 16$-supermatrix
which belongs to the coset-space defining the non-linear $\sigma$-model
\cite{verbaarschot,efetov1,altland1} for the orthogonal symmetry class.

Now, we consider the limit $M\gg 1$ and perform an expansion in
terms of the small parameter $x=(1+M)^{-1}$.
In typical experiments with ballistic quantum dots,
the number of channels is about $3$-$6$ and correspondingly
$x=1/7$-$1/13$. Using the well known square root parametrization
\cite{verbaarschot,iida1}, the $Q$-matrix can be written as
$Q = T^{-1}\Lambda T$, $\ T=\sqrt{1+D^2}+D$.
$D$ is a supermatrix which has only non diagonal entries in the
$\Lambda$-grading: $D_{11}=D_{22}=0$ and $D_{21}=L_g\,D_{12}$.
The integration measure \cite{iida1} is
$dQ=\exp\left(-{\textstyle \frac{1}{4}\mbox{Str}\ \ln(1+D^2)}\right)
d\mu(D_{12})$
where $d\mu(D_{12})$ is the flat measure of the $8\times 8$ supermatrix
$D_{12}$. The Jacobian factor of this measure modifies the action
in Eq. (\ref{act_def}), i.~e. in the first contribution $M$ is
be replaced by $M+1$. The $Q$-integral can then be evaluated perturbatively
\cite{iida1} where the (modified) action has to be
expanded in even powers of $D$, ${\cal L}(D)={\cal L}_2(D)+{\cal L}_4(D)+
{\cal L}_6(D)+\cdots$. The expansion of the source term contributions
and of the exponential of ${\cal L}_4(D)+{\cal L}_6(D)$ up to three leading
orders in $x$ yields a Gaussian integral with the action ${\cal L}_2(D)$.
The supermatrix $D_{12}$ has to be decomposed into the 16 blocks
corresponding to the $\tau_3$- and $\Sigma_3$-grading. Each of these blocks
is associated with a particular propagator determined by ${\cal L}_2(D)$.
Due to the large number of different terms in the higher order supertraces,
we applied a computer program, which was developed for this purpose in C++,
for the final evaluation of the integral. To get the autocorrelation,
one has to subtract from (\ref{sigma_int1}) the product of the two
independently averaged conductance contributions which have been
calculated similarly.

Before we discuss the final result, we mention that the
action (\ref{act_def}) also describes
diffusive metallic systems \cite{efetov1,altland1} provided that all
relevant energy scales, in particular the level broadening $M\cdot\Delta_0$,
are smaller than the Thouless energy. The perturbation theory applied here
corresponds exactly \cite{efetov1}
to the standard diagrammatic approach in terms of diffuson and cooperon
modes. The relevant contributions for the conductance autocorrelation
are the diagrams \cite{lee} consisting of two connected conductance
loops.
In the $\sigma$-model, they correspond \cite{iida1,amg,altland2}
to the contributions of $D_{12}$ that are non diagonal in the
$\Sigma_3$-grading and are associated with
two propagators: the cooperon mode
$P_C=[(M+1)(1+\varphi^2+\frac{\kappa^2}{4}\Delta\varphi^2+i\Delta\eps)]^{-1}$
(non diagonal in the $\tau_3$-grading) and the diffuson mode
$P_D=[(M+1)(1+\frac{1+\kappa^2}{4}\Delta\varphi^2+i\Delta\eps)]^{-1}$
(diagonal in the $\tau_3$-grading). The conductance autocorrelation
is estimated \cite{lee} as $\sim M^2(|P_D|^2+|P_C|^2)$ which is indeed
confirmed by the explicit computer calculation. We obtain (after
the shifts $\varphi\to\varphi+\frac{1}{2}\Delta\varphi$,
$\eps\to\eps+\frac{1}{2}\Delta\eps$)
for the average conductance $\langle g(\varphi,\,\eps)\rangle$ and the
autocorrelation function
$C_{\varphi,\eps}(\Delta\varphi,\,\Delta \eps)=\langle \delta
g(\varphi,\,\eps) \,\delta g(\varphi+\Delta\varphi,\,\eps+\Delta
\eps)\rangle$\ \ ($\delta g=g-\langle g\rangle$)
the results
\begin{eqnarray}
\label{g_mean}
\langle g(\varphi,\eps)\rangle & = & \frac{M}{4}-\frac{M}{4(M+1)}\,
\frac{1}{1+\varphi^2}\quad,\\
\label{g_auto}
C_{\varphi,\eps}(\Delta\varphi,\Delta\eps)&=&
\frac{1}{16}\Biggl(\frac{1}{\left(1+(\varphi+\frac{1}{2}\Delta\varphi)^2
+\frac{\kappa^2}{4}\Delta\varphi^2\right)^2+\Delta\eps^2}
+\frac{1}{\left(1+\frac{1+\kappa^2}{4}\Delta\varphi^2\right)^2
+\Delta\eps^2}\Biggr)\ ,
\end{eqnarray}
which are correct up to terms of order ${\cal O}(x^2)$ for
(\ref{g_mean}) and of order ${\cal O}(x)$ for (\ref{g_auto}).

The quantities $\varphi=(8N\alpha^2/(M+1))^{1/2}$,
and $\Delta\eps=2\pi E/[(M+1)\Delta_0]$, given
in terms of the original model parameters,
are of course related to the corresponding
physical quantities in a ballistic quantum dot. The energy correlations
decay on a scale $\gamma=\Delta_0(M+1)/\pi$ which is just the level
broadening (or inverse life time) due to the coupling with the channels
\cite{iida1}.
$\Delta_0$ has to be adjusted to the level spacing in the quantum dot
at the Fermi energy. The relation between the parameter $\sqrt{N}\alpha$
and the magnetic flux $\Phi$ applied on the dot was studied in
\cite{bohigas5,frahm1} and is given by
$\sqrt{N}\,\alpha=const.\,\sqrt{\hbar v_F/(L\Delta_0)}\,(\Phi/\Phi_0)$ where
$v_F$ is the Fermi velocity, $L$ a typical
diameter of the quantum dot and $\Phi_0=h/e$
the flux quantum. The numerical constant is of the order of unity and
depends on geometrical details. For a particular model, a circle (radius $L$)
with a very rough surface, it takes the value $\sqrt{4/3}\simeq 1.15$
\cite{frahm1}. To establish the connection with experiment one should
therefore set
$\varphi=const.\,(8\hbar v_F/L\gamma)^{1/2}\,(\Phi/\Phi_0)$
and $\Delta\eps=\Delta E/(2\gamma)$ in Eqs. (\ref{g_mean}) and
(\ref{g_auto}).
The scale dependence on the channel number $M$ (via $\gamma$) can be
understood in terms of semiclassical trajectories. With increasing
number of channels (decreasing life time $\hbar/\gamma$) the typical length
of the trajectories (and therefore the enclosed area) decreases so that the
critical value for the energy (or the magnetic flux) to break the
phase-coherence of the electrons is increased.

In the limiting cases of orthogonal ($\beta=1$, $\varphi=0$) and
unitary ($\beta=2$, $\varphi\to\infty$) symmetry, the correlation function
becomes a squared Lorentzian in $\Delta\varphi$ and a simple Lorentzian
in $\Delta\eps$
\begin{equation}
\label{g_auto_limit}
C_{\varphi,\eps}(\Delta\varphi,\Delta\eps)=
\frac{1}{8\beta}\,
\frac{1}{\left(1+\frac{1+\kappa^2}{4}\Delta\varphi^2\right)^2+\Delta\eps^2}
\quad.
\end{equation}
For $\kappa=0$, $\beta=2$ this expression is consistent
with the semiclassical approach \cite{jalabert3}
and the result of Ref. \cite{efetov2}, provided the scales of
the flux are properly translated. The prefactor in (\ref{g_auto_limit})
exhibits the $\beta$-dependence of the universal conductance fluctuations
for a ballistic cavity obtained by the random matrix approach for
the scattering matrix \cite{jalabert1,baranger1}.
The expression (\ref{g_mean}) for the average conductance
is included here for completeness and agrees
very well with the (numerical fit of the) precise result of
Ref. \cite{pluhar}.

The semiclassical calculations of Refs. \cite{jalabert3,jalabert4}
do not reproduce the correct amplitude but they predict the same functional
dependence of the weak localization peak and the autocorrelation
function on the magnetic field as (\ref{g_mean}) and (\ref{g_auto_limit}),
provided the imaginary antisymmetric symmetry breaking ($\kappa=0$)
is considered. We also recover the simple Lorentzian for the energy
correlation \cite{bluemel1}.

In Refs. \cite{jalabert1,baranger1} the $S$-matrix of the quantum dot
was described by one of the circular ensembles \cite{dyson1}
for orthogonal or unitary symmetry. The crossover between these
cases can be modelled by a so-called
Brownian motion ensemble \cite{frahm1,rau1} where the $S$-distribution
depends on a fictitious time $t$ corresponding for $t=0$ to the
circular orthogonal ensemble and ``diffusing'' for $t\to\infty$ to
the circular unitary ensemble as stationary distribution.
The average conductance and the autocorrelation function
obtained in this approach are given by \cite{frahm1,rau1}:
\begin{equation}
\label{brown_g_mean}
\langle g(t)\rangle = \frac{M}{4}-\frac{M}{4(M+1)}\,
e^{-t/t_c}\quad,\quad
\langle \delta g(t)\,\delta g(t+\Delta t) \rangle
=\frac{1}{16}\,e^{-\Delta t/t_c}\,\left(1+e^{-2t/t_c}\right)
+{\cal O}(M^{-1})
\end{equation}
where $t_c$ is the critical time that determines the crossover scale.
The comparison of the average conductance expressions
(\ref{brown_g_mean}) and (\ref{g_mean}) suggests the
identification $t=t_c\ln(1+\varphi^2)$ between the Brownian
motion time $t$ and the parameter $\varphi$. Then, the
conductance fluctuations given by (\ref{g_auto})
($\Delta\varphi=\Delta\eps=0$) and (\ref{brown_g_mean}) ($\Delta t=0$)
coincide. Concerning the correlations, we obtain
from (\ref{g_auto_limit}) (for the limiting cases $\varphi=0$
or $\varphi\to\infty$) in lowest order in $\Delta\varphi^2$:
$C_{\varphi,\eps}(\Delta\varphi,0)\simeq\frac{1}{8\beta}
(1-\frac{1}{2}(1+\kappa^2)\Delta\varphi^2)$, which coincides with the
Brownian motion expression (with $\Delta t/t_c\simeq\Delta\varphi^2$)
only for the case of hermitian symmetry breaking ($\kappa=1$).
It is indeed possible \cite{frahm1} to relate the Brownian motion
approach directly (for small $\varphi$, $\Delta\varphi$, large $M$)
to the model considered here ($\kappa=1$), giving
the correct identification $\Delta\varphi^2=\Delta t/t_c$. In the
crossover regime $\varphi\sim 1$, this identification and the
mapping onto the Brownian motion approach do not work
since the correlation function (\ref{g_auto}) is not even
in $\Delta\varphi$.

In summary, we have calculated the conductance autocorrelation function
(with respect to magnetic field and energy) of a chaotic quantum dot
for the complete crossover from orthogonal to unitary symmetry.
The perturbative approach seems to work rather well, since our expression
(\ref{g_mean}) for the weak localization peak agrees very well
with the precise result of Ref. \cite{pluhar}. In addition, we have found
that different
types of symmetry breaking lead to different scales in the autocorrelation
function
explaining the inconsistencies between semiclassical theory
($\kappa=0$) and the Brownian motion approach ($\kappa=1$).

\begin{displaymath}
***
\end{displaymath}

The author thanks J.-L. Pichard, A. M\"uller-Groeling, H. A. Weidenm\"uller,
and C. W. J. Beenakker for helpful discussions and acknowledges the D.F.G.
for a post-doctoral fellowship.

\end{document}